\documentstyle[11pt,newpasp,twoside,epsf]{article}
\def\edcomment#1{\iffalse\marginpar{\raggedright\sl#1\/}\else\relax\fi}
\reversemarginpar

\begin{document}
\title{The Origin of Hot Subluminous Horizontal-Branch Stars
in $\omega$~Centauri and NGC~2808}
\author{Allen V. Sweigart}
\affil{NASA Goddard Space Flight Center, Code 681, Greenbelt, MD 20771}
\author{Thomas M. Brown}
\affil{Space Telescope Science Institute, 3700 San Martin Drive,
Baltimore, MD 21218}
\author{Thierry Lanz, Wayne B. Landsman, Ivan Hubeny}
\affil{NASA Goddard Space Flight Center, Code 681, Greenbelt, MD 20771}

\begin{abstract}
Hot subluminous stars lying up to 0.7 mag below the
extreme horizontal branch (EHB) are found in the
ultraviolet (UV) color-magnitude diagrams of both
$\omega$~Cen (D'Cruz et al. 2000) and NGC~2808 (Brown et al.
2001).  In order to explore the evolutionary status of these
subluminous stars, we have evolved a set of
low-mass stars continuously from the main sequence
through the helium-core flash to the HB for a wide range
in the mass loss along the red-giant branch (RGB).  Stars
with the largest mass loss evolve off the RGB to high
effective temperatures before igniting helium in their
cores.  Our results
indicate that the subluminous EHB stars, as well as the
gap within the EHB of NGC~2808, can be
explained if these stars undergo a late helium-core flash
while descending the white-dwarf cooling curve.  Under
these conditions the convection zone produced by the
helium flash will penetrate into the stellar envelope,
thereby mixing most, if not all, of the envelope hydrogen
into the hot helium-burning interior, where it is rapidly
consumed (Sweigart 1997).  This phenomenon is analogous to
the ``born-again'' scenario for producing hydrogen-deficient
stars following a very late helium-shell flash.
This ``flash mixing'' of the stellar envelope
greatly enhances the envelope helium
and carbon abundances and, as a result, leads to a
discontinuous jump in the HB effective temperature.
We argue that the EHB gap in NGC~2808 is
associated with this theoretically predicted dichotomy in
the HB morphology.  Using new helium- and carbon-rich
stellar atmospheres, we show that these changes in the
envelope abundances of the flash-mixed stars will suppress
the UV flux by the amount needed to explain the
hot subluminous EHB stars in $\omega$~Cen and NGC~2808.
Moreover, we demonstrate that models without flash mixing
lie, at most, only $\sim 0.1$ mag below the EHB, and
hence fail to explain the observations.  Flash mixing
may also provide a new evolutionary channel
for producing the high gravity, helium-rich sdO and sdB stars.
\end{abstract}

\section{Introduction}

Extreme horizontal-branch (EHB) stars occupy the hot end
of the horizontal branch (HB) in globular clusters with
extended blue HB tails.  The envelope masses of these
stars are too small ($<$~$\sim$0.02~$M_\odot$) to sustain
hydrogen-shell burning, and thus nearly all of their
luminosity comes from helium burning in the core.  Recent
observations have discovered an unexpected population
of hot stars lying below the canonical EHB
in the UV color-magnitude diagram (CMD) of
$\omega$~Cen (D'Cruz et al. 2000 and references
therein).  In the present paper we will use new UV observations of
the globular cluster NGC~2808 to explore the origin of
these subluminous EHB stars and will suggest that these
stars may be the progeny of stars which underwent
extensive mixing during a delayed helium flash on the white-dwarf
(WD) cooling curve (Brown et al. 2001).

Our data for NGC~2808 were obtained
in the far-UV (FUV, $\lambda \sim 1600$~\AA) and near-UV
(NUV, $\lambda \sim 2700$~\AA) bandpasses
of the Space Telescope Imaging
Spectrograph (STIS).  The HB of
NGC~2808 is bimodal with a large gap between the
blue HB (BHB) and red HB (RHB) stars.  In
addition, NGC~2808 has a very long blue HB tail that is
punctuated by two gaps: one between the EHB and BHB and
one within the EHB itself (Sosin et al. 1997; Walker
1999; Bedin et al. 2000).  Our
STIS CMD (Figure 1) shows the following features:

\begin{itemize}
  \item The gap between the EHB and BHB
at $m_{FUV}-m_{NUV} \sim -1$~mag is well
detected, as is the gap between
the BHB and RHB at $m_{FUV}-m_{NUV} > 0$~mag.  The
gap within the EHB seen in optical CMDs is
not present.
  \item There is a large population of hot subluminous HB
stars, previously known to exist only in $\omega$~Cen
(D'Cruz et al. 2000).  Out of a total of 75 EHB stars, 46
are fainter than the canonical zero-age horizontal branch
(ZAHB).
  \item 5 luminous post-EHB stars are found
at $m_{FUV} < 15.5$~mag.
\end{itemize}

\section{Canonical Evolution through the Helium Flash}

To study the origin of the hot subluminous EHB stars, we
computed a set of evolutionary sequences which followed
the evolution of a low-mass star continuously from the
main sequence through the helium flash to the
ZAHB.  All sequences had the same heavy-element
abundance Z of 0.0015, corresponding to [Fe/H] = $-$1.31
for [$\alpha$/Fe] = 0.3, and an initial main-sequence
mass $M$ of $0.862~M_\odot$, corresponding to an age
of 13 Gyr at the tip of the RGB.  The sequences only
differed in the extent of mass loss along the RGB
which we specified by varying the mass-loss
parameter $\eta_R$ in the Reimers formulation
from 0 to 1.  Our purpose was to determine
if mixing between the
helium core and hydrogen envelope might occur
during the helium flash, as found previously
by Sweigart (1997).

A representative sample of our evolutionary tracks is shown
in Figure 2.  For $0.0 \le \eta_R < 0.740$ the helium flash
occurs at the tip of the RGB.  As the mass loss increases,
however, the models peel off the RGB and evolve to high
effective temperatures before igniting helium
(so-called ``hot He-flashers''; Castellani \& Castellani
1993; D'Cruz et al. 1996).  For $0.740 \le \eta_R \le 0.817$ the
helium flash occurs between the tip of the RGB and the top of the
WD cooling curve.

\vspace{\fill}
\begin{figure*}[t]
\hspace{-0.50in}{\epsfbox{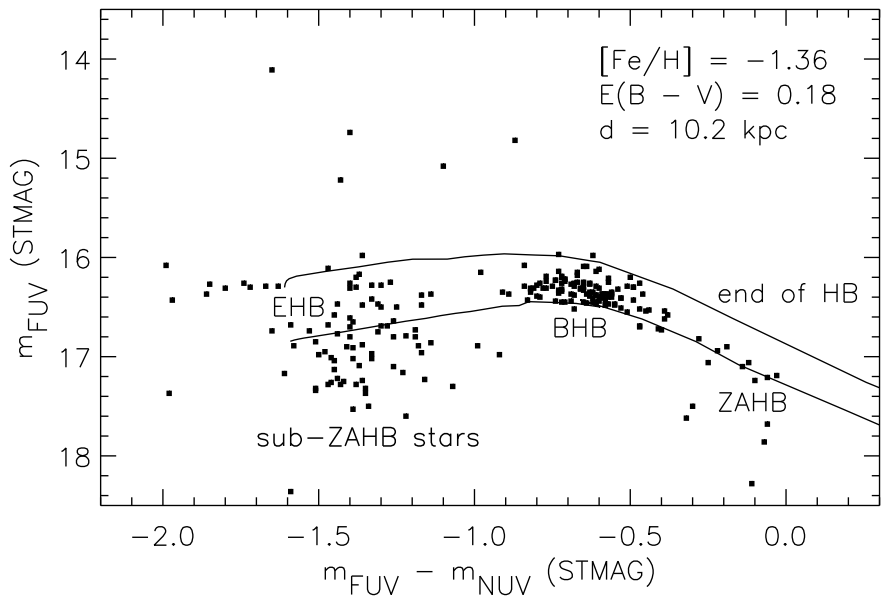}}
\vspace{-0.10in}
\caption{STIS UV color-magnitude diagram for NGC~2808.
The theoretical loci for the ZAHB and the end of the HB
phase (solid lines) were transformed to the observational plane
using the indicated cluster parameters.  Note the large
population of hot sub-ZAHB stars.}
\label{figure 1}
\end{figure*}
\vspace{-0.10in}

In all sequences with $\eta_R \le 0.817$ the flash
convection zone produced by the high helium-burning
luminosity (peak $L_{He} \sim 10^{10}~L_\odot$)
failed to reach the hydrogen envelope
by $\sim$1 pressure-scale height.  Thus
mixing between the helium
core and the hydrogen envelope does not occur if
a star ignites helium either on the RGB
or during the evolution to the top of the WD
cooling curve.  In these cases we confirm the
canonical assumption that the helium flash does not
affect either the envelope mass or composition.

\section{Flash Mixing on the White-Dwarf Cooling Curve}

\begin{figure*}[!ht]
\hspace{-0.55in}{\epsfbox{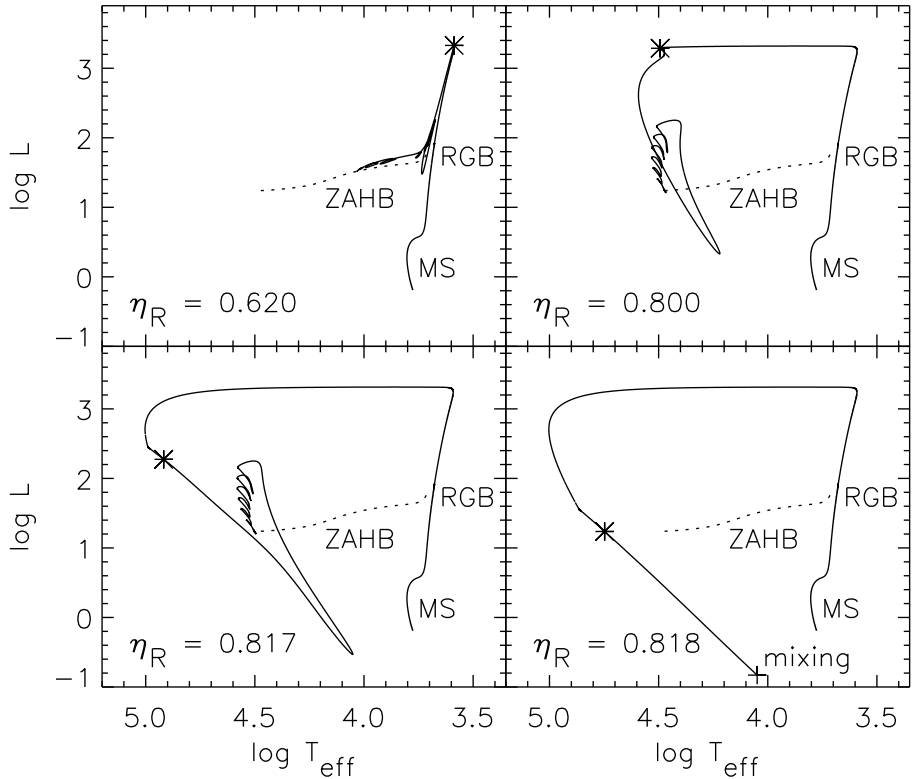}}
\vspace{-0.15in}
\caption{Evolutionary tracks from the main
sequence (MS) through the helium flash to the ZAHB for 4
values of the mass-loss parameter $\eta_R$.  The
peak of the helium flash (*) is indicated.  The
flash convection reached the hydrogen
envelope at the end of the $\eta_R$ = 0.818 track
(+).}
\label{figure 2}
\end{figure*}
\vspace{-0.10in}

The canonical evolution described above changes dramatically when
the helium flash occurs further down the WD
cooling curve (Figure 2).  As a star descends the
cooling curve, the entropy barrier of its
hydrogen shell decreases (Iben 1976).  As a result, the
flash convection is then able to penetrate deeply into the hydrogen
envelope (Sweigart 1997; Brown et al. 2001).  The
protons captured by the flash convection will be mixed
into the helium-burning core
while helium and carbon from the core will be
mixed outward into the envelope.  The calculations
of Sweigart (1997) indicate that this ``flash mixing''
will consume most, if not all, of the envelope
hydrogen while simultaneously
enriching the envelope in helium and carbon.  All
of our sequences with $0.818 \le \eta_R \le 0.936$
encountered flash mixing.  These sequences were stopped
at the onset of mixing due to the numerical difficulty of
following the proton mixing and nucleosynthesis.
Sequences with $\eta_R \geq 0.937$ did not ignite helium
and thus died as helium white dwarfs.

Flash mixing is a consequence of the basic properties of
the stellar models and hence should occur
whenever a star ignites helium on the
WD cooling curve.  Analogous mixing occurs
during a very late helium-shell flash
according to the ``born-again'' scenario for producing
hydrogen-deficient stars (Iben 1995 and references
therein).

EHB evolutionary tracks for both canonical ($\eta_R \le 0.817$)
and flash-mixed ($0.818 \le \eta_R \le 0.936$)
sequences are plotted in Figure 3.  The canonical (i.e., unmixed)
models have the same H-rich
envelope composition as the pre-helium flash models. The
ZAHB models for the flash-mixed tracks were obtained by
changing the envelope composition
to mimic the effects of
flash mixing.  We considered both He + C-rich and He-rich
envelope compositions for these flash-mixed models (see caption
of Figure 3).  For comparison the bottom panel of Figure 3
shows a set of tracks with $0.818 \le \eta_R \le 0.936$ which
have the same H-rich envelope composition as the canonical
models.

\begin{figure*}[p]
\hspace{-0.45in}{\epsfbox{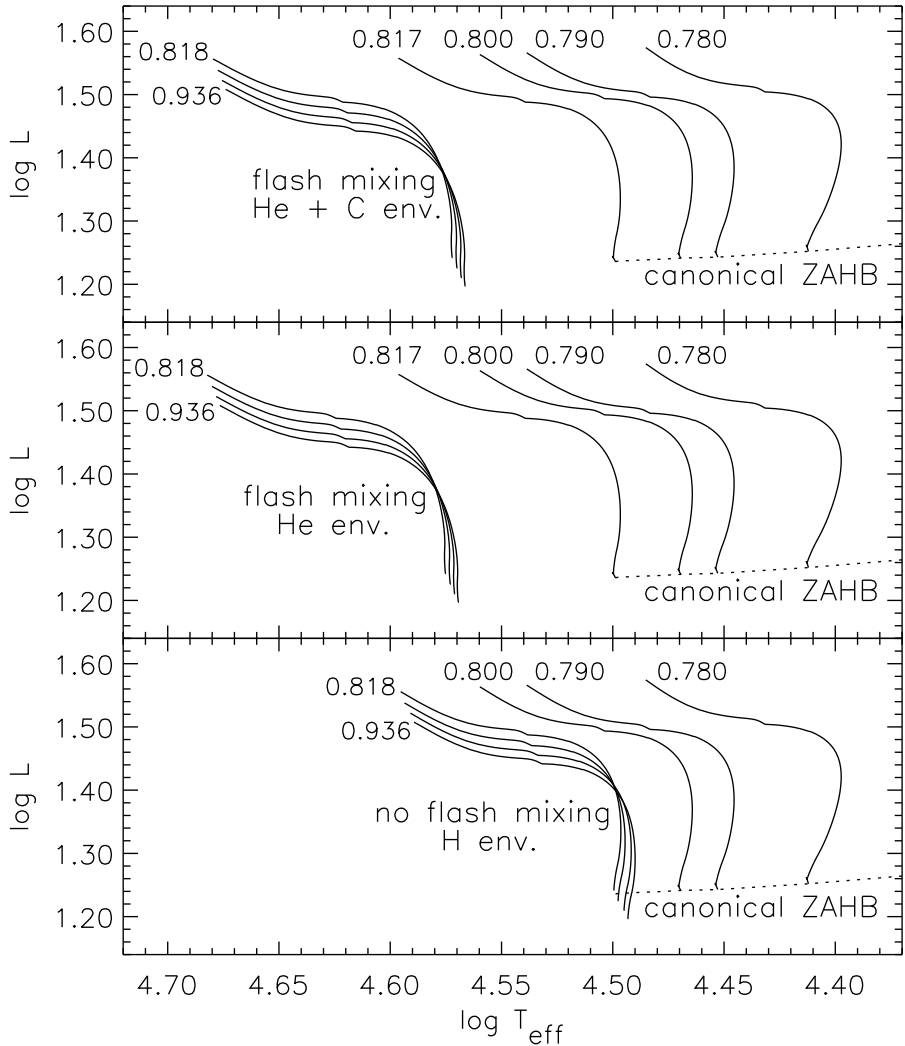}}
\vspace{-0.10in}
\caption{EHB evolutionary tracks for canonical
sequences ($\eta_R \leq 0.817$) and sequences with flash
mixing ($\eta_R$ = 0.818, 0.860, 0.900, 0.936).  The ends of
the tracks are labeled by the value of $\eta_R$.  The
flash-mixed tracks in the top panel have an
envelope helium abundance Y of 0.96 and a carbon
abundance C of 0.04 by mass, with the other heavy
elements at their cluster abundances.  The flash-mixed
tracks in the middle panel are the same except
that Y = 1.0 and C = 0.0.  The corresponding tracks in
the bottom panel have the same H-rich envelope
composition as the canonical tracks.  Note the
temperature gap between the He + C and He
flash-mixed tracks and the canonical tracks.  In
contrast, the H tracks, which ignore flash mixing, lie at
the hot end of the canonical EHB.}
\label{figure 3}
\end{figure*}

Several features of Figure 3 deserve comment.  First,
there is a well-defined high temperature limit to the
canonical EHB at an effective temperature $\rm T_{eff}$ of
$\sim$31,500~K for the present metallicity.  Within the
canonical framework it is not possible to produce hotter
EHB stars regardless of the extent of mass loss along the
RGB.  Second, the flash-mixed tracks are separated by a gap of
$\sim$6000~K in $\rm T_{eff}$ from the hot end of the
canonical EHB.  As a consequence,
flash mixing leads to a sharp dichotomy in the
properties of the EHB stars.  The change from canonical
to flash-mixed evolution occurs over an interval of only
0.001 in $\eta_R$, corresponding to a mass-loss
difference of only 10$^{-4}~M_\odot$.

\section{Nature of the Subluminous EHB Stars}

In order to compare our EHB models with the STIS data,
we computed new stellar atmospheres for both H-rich
and H-depleted compositions.  The spectral
energy distribution for a canonical EHB star is
compared to the spectra for 2 flash-mixed stars with He
+ C-rich and He-rich atmospheres in Figure 4.  The
enhanced helium abundance in the flash-mixed stars
reduces the hydrogen opacity below 912~\AA~so that more of
the flux is radiated in the extreme UV at the expense of
the flux at longer wavelengths.  Enhancing the carbon
abundance along with helium restores some of this extreme
UV opacity, but the resulting spectrum is still redder
and fainter in the far-UV than a normal spectrum.

\begin{figure*}[!h]
\hspace{-0.38in}{\epsfbox{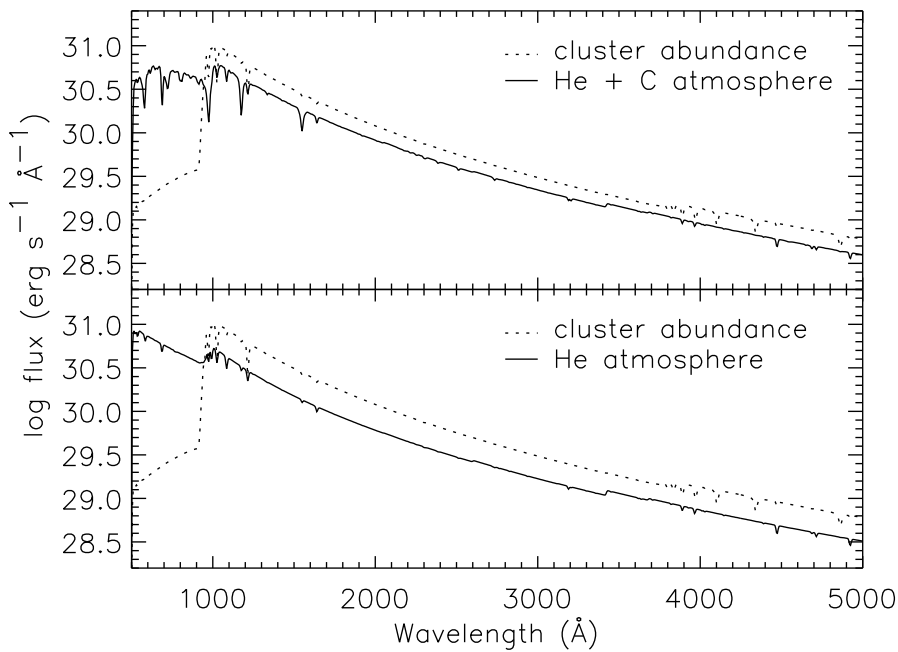}}
\vspace{-0.10in}
\caption{Spectral energy distribution for a canonical
EHB star ($\rm T_{eff} = 31,000$~K, dashed lines) with the nominal
cluster abundance of NGC~2808 and for two flash-mixed stars
($\rm T_{eff} = 37,000$~K, solid lines) with either a He + C
(upper panel) or He (lower panel) envelope
composition.  Note that the flash-mixed stars
are fainter in the UV.}
\label{figure 4}
\end{figure*}

Using these He + C-rich and He-rich stellar atmospheres,
we translated the ZAHB models from the flash-mixed
sequences to the STIS CMD (see Figure 5).  The
flash-mixed models show the same dramatic drop in the far-UV
luminosity as the STIS data.  This drop can be attributed
to 1) the lower far-UV flux in the spectral energy
distribution of the flash-mixed models (see Figure 4)
and 2) the larger bolometric correction due to
their higher temperatures (see Figure 3).  Evolution off
the ZAHB will fill in the region between these flash-mixed
models in Figure 5 and the canonical ZAHB.  Radiative
levitation of Fe would shift the flash-mixed
models redward, possibly explaining the color spread of
the subluminous stars.  In contrast,
the $\eta_R$ = 0.860 ZAHB model with a normal H-rich envelope
in Figure 5 lies just $\sim$0.1~mag fainter than the hot end
of the canonical EHB, a consequence
of its slightly smaller core mass ($\Delta M_c \sim 0.004~M_\odot$).  It
is clear therefore that
hot He flashers which ignore
flash mixing cannot explain the subluminous stars in
either $\omega$~Cen or NGC~2808.

\begin{figure*}[!h]
\hspace{-0.42in}{\epsfbox{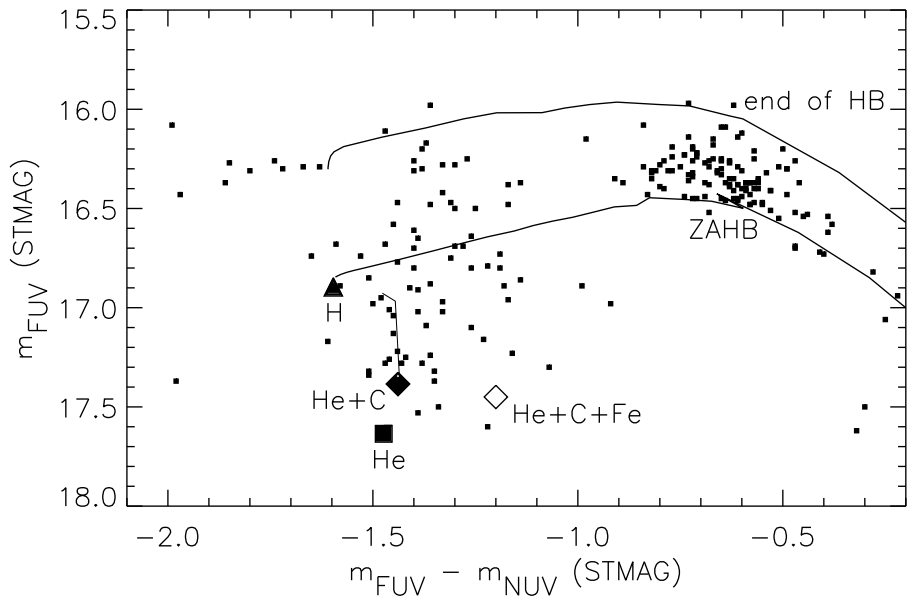}}
\vspace{-0.15in}
\caption{Location of the $\eta_R$ = 0.860 ZAHB models
in the STIS CMD of NGC~2808.  The ZAHB model with a normal H-rich
envelope (filled $\bigtriangleup$) lies near the hot
end of the canonical EHB and thus cannot explain the
subluminous stars.  However, the flash-mixed ZAHB models
with He + C (filled $\Diamond$) or He (filled $\Box$)
envelopes lie well below the canonical EHB.  Such models
evolve to brighter luminosities (thin line), filling the
area under the canonical EHB.  Radiative levitation
of iron to [Fe/H] = +1 would move
the He + C model redward (open $\Diamond$).}
\label{figure 5}
\end{figure*}

Optical CMDs also show a gap within the EHB
of NGC~2808.  We have plotted
our ZAHB models in the ($B-V$, $V$) CMD
to see if the temperature difference between the flash-mixed
and canonical tracks in Figure 3 might produce such
a gap.  As expected, the flash-mixed models are separated by a gap of
$\sim$0.5~mag in V from the hot end of the canonical
EHB.  Our simulations, which include evolution
off the ZAHB, predict an EHB gap close to that observed
in NGC~2808 (see Brown et al. 2001).  A similar EHB gap
is not seen in $\omega$~Cen, possibly because it has
been obscured by the metallicity spread of the cluster.

\section{Conclusions}

\begin{itemize}
  \item Stars which ignite helium on the white-dwarf cooling
curve will undergo substantial mixing between the helium
core and hydrogen envelope during the helium flash.
This flash-induced mixing will greatly enhance the
envelope helium and carbon abundances.
  \item Flash-mixed EHB stars will appear subluminous in UV
color-magnitude diagrams due to their redistributed far-UV flux
and larger bolometric corrections, in agreement with the
subluminous stars in $\omega$~Cen and NGC~2808.
  \item Flash mixing leads to a dichotomy in the
properties of EHB stars.  The temperature gap between the
flash-mixed and canonical (i.e., unmixed) stars may
explain the EHB gap found in optical CMDs of NGC~2808.
  \item Flash mixing on the white-dwarf cooling curve may provide
a new evolutionary channel for producing H-deficient
stars, particularly the high gravity, He-rich sdO stars
(Lemke et al. 1997) and the minority of sdB stars that
are He-rich (Moehler, Heber, \& Durrell 1997).
\end{itemize}

\end{document}